\documentclass[letterpaper]{article} 
\usepackage{aaai25}  
\usepackage{times}  
\usepackage{helvet}  
\usepackage{courier}  
\usepackage[hyphens]{url}  
\usepackage{graphicx} 
\usepackage{natbib}  
\usepackage{caption} 
\usepackage{amssymb}
\usepackage{algorithm}
\usepackage{algorithmic}
\usepackage{amsfonts,amssymb}
\usepackage{multirow}
\usepackage{multicol}
\usepackage{subfigure}
\usepackage{booktabs}
\usepackage{aligned-overset}
\usepackage{newfloat}
\usepackage{listings}
\DeclareCaptionStyle{ruled}{labelfont=normalfont,labelsep=colon,strut=off} 
\floatstyle{ruled}
\newfloat{listing}{tb}{lst}{}
\floatname{listing}{Listing}
\title{Leveraging Contrastive Learning and Self-Training for Multimodal Emotion Recognition with Limited Labeled Samples}
\author{
    Qi Fan \textsuperscript{\rm 1}\equalcontrib,
    Yutong Li \textsuperscript{\rm 2}\equalcontrib,
    Yi Xin \textsuperscript{\rm 3}\thanks{Corresponding Author.},
    Xinyu Cheng \textsuperscript{\rm 3},
    Guanglai Gao \textsuperscript{\rm 1},
    Miao Ma \textsuperscript{\rm 2}\\
}
\affiliations{
    ~
    \\   
    \textsuperscript{\rm 1}Inner Mongolia University,~~
    \textsuperscript{\rm 2}Shaanxi Normal University,~~
    \textsuperscript{\rm 3}Nanjing University \\
    ~
    \\
    fanqi1203@foxmail.com, liyutongstu@snnu.edu.cn, xinyi@smail.nju.edu.cn, chengxinyu@smail.nju.edu.cn\\
    csggl@imu.edu.cn, mmthp@snnu.edu.cn
}

\usepackage{bibentry}

\begin{document}

\maketitle

\begin{abstract}
The Multimodal Emotion Recognition challenge MER2024 focuses on recognizing emotions using audio, language, and visual signals. In this paper, we present our submission solutions for the Semi-Supervised Learning Sub-Challenge (MER2024-SEMI), which tackles the issue of limited annotated data in emotion recognition. Firstly, to address the class imbalance, we adopt an oversampling strategy. Secondly, we propose a modality representation combinatorial contrastive learning (MR-CCL) framework on the trimodal input data to establish robust initial models. Thirdly, we explore a self-training approach to expand the training set. Finally, we enhance prediction robustness through a multi-classifier weighted soft voting strategy. Our proposed method is validated to be effective on the MER2024-SEMI Challenge, achieving a weighted average F-score of 88.25\% and ranking 6th on the leaderboard. Our project is available at https://github.com/WooyoohL/MER2024-SEMI.
\end{abstract}

\section{Introduction}
Multimodal Emotion Recognition (MER) has increasingly garnered interest due to its potential in various applications including human-computer interaction (HCI)~\cite{moin2023emotion,li2024mie, song23c_interspeech}, mental health monitoring~\cite{xu2022measurement,Shen2024}, and social media analysis~\cite{chandrasekaran2021multimodal, yu2024credit}. Research in MER primarily concentrates on three modalities: text, audio, and visual. There has been substantial progress in this field with the adoption of advanced techniques such as deep learning and multimodal fusion. Recent studies, for instance, Li et al.  ~\cite{li2021retracted} have introduced architectures like Deep Convolutional Neural Networks (DCNN) and Deep Separable Convolutional Neural Networks (DSCNN) for speech and face recognition. MFN~\cite{8925497} explores how attention mechanisms can be successfully applied to model emotion recognition from rich narrative videos. DEFR~\cite{zhao2021former} incorporates a novel transformer approach that combines convolutional spatial transformers with temporal transformers for enhanced extraction of facial features over time and space. Despite these advancements, challenges remain, particularly in
the time-consuming and costly process of sentiment data annotation. The reliance on extensively annotated datasets severely restricts the practical applicability of most existing multimodal sentiment recognition methods.

In response to the above problems, the ACM MM unveiled the MER2024-SEMI challenge. This initiative requires participants to predict discrete emotions with a limited set of labeled data, complemented by a significant volume of unlabeled data for training. 
To enhance model generalization, the challenge encourages the exploration of large-scale unlabeled data utilization. One effective strategy is semi-supervised learning (SSL), particularly self-training methods that have shown promise in various domains~\cite{liu2024enhanced, xin2023self, Li-ICMR1}. These methods generate pseudo-labels from unlabeled data, integrating these with labeled data during the training process. Another promising strategy involves self-supervision~\cite{chen2020simple, chen2020improved, liu2024contrastive}, which entails pretraining models on large amounts of unlabeled data and then fine-tuning them for specific tasks~\cite{xin2024vmt, xin2024mmap}. In the context of the MER2024-SEMI challenge, participants must contend with a discrete set of six emotions: neutral, anger, happiness, sadness, worry, and surprise. Importantly, the distribution of data across these classes varies (as shown in Figure~\ref{fig1}), posing additional challenges related to class imbalance that participants must address.

To tackle the MER2024-SEMI challenge, we propose a straightforward yet robust semi-supervised multimodal emotion recognition method. To address the class imbalance, we implement a crucial oversampling strategy, focusing on expanding classes with limited data. Next, we propose a modality representation combinatorial contrastive learning framework for effectively utilizing the provided unlabeled data in the trimodal input. Additionally, we explore semi-supervised techniques, particularly finding the self-training strategy valuable. This approach involves generating pseudo-labels for unlabeled data and integrating them into the training process. Furthermore, to enhance the robustness of the prediction results, we employ ensemble learning via aggregating the confidence scores of multi-classifiers. Our approach achieves an 88.25\% weighted average F-score on MER2024-SEMI.

\section{Method}

\subsection{Problem Formulation}
MER2024-SEMI challenge provides a small amount of labeled data ${D_l} = \{ ({x_i},{y_i})\} _{i = 1}^{{N_l}}$ and and a large amount of unlabeled data ${\tilde D_u} = \{ ({\tilde x_i})\} _{i = 1}^{{N_u}}$, where ${y_i}$ is the discrete label of the sample ${x_i}$ and ${y_i} \in \{ 1,2,...,C\}$. Here, ${N_l}$, ${N_u}$, and $C$ respectively denote the number of labeled samples, unlabeled samples, and emotion classes. For each ${x_i}$ (or ${\tilde x_i}$), we extract visual features $F_i^v$, acoustic features $F_i^a$ and text features $F_i^t$ from the visual, audio, and text modalities, where $F_i^v \in \mathbb{R}^{d_v}$, $F_i^a \in \mathbb{R}^{d_a}$, $F_i^t \in \mathbb{R}^{d_t}$, and ${\{ {d_m}\} _{m \in \{ v,a,t\} }}$ represent the feature dimensions for each modality. For simplicity, we omit the subscript $i$ in the following when there is no ambiguity.

\subsection{Feature Extraction}\ 

For the whole dataset, we extract the utterance level feature according to the MER2024 baseline \cite{lian2023mer, 2024mer}.

\textbf{Visual feature:} Since the initial samples in MER2024 are in video format, we first crop and align the facial regions of each frame using the OpenFace toolkit
\cite{baltruvsaitis2016openface}. Subsequently, we leverage the pre-trained CLIP-Large model
\cite{radford2021learning} to extract frame-level features for each face image. These frame-level visual features are then aggregated using average pooling to generate video-level embeddings.

\textbf{Acoustic feature:} First, we utilize the FFmpeg toolkit 
to separate the audio from the video with a sampling rate of 16kHz. Following this, we employ the Chinese-HuBert-Large acoustic encoder 
\cite{hsu2021hubert} to extract features, which performs well on Chinese sentiment corpus. The averaged hidden representations from the last 4 layers of the model are used as the final acoustic representations due to their high sensitivity to the semantic information. 

\textbf{Text feature:} We first convert audio files into transcripts by employing WeNet 
\cite{yao2021wenet}, an open-source automatic speech recognition toolkit. Then, we choose the Baichuan2-13B-Base model  
\cite{baichuan2023baichuan2} that has been pre-trained on a large-scale corpus as a feature extractor and further input transcripts into the language model to obtain a feature vector as the representation of the text modality.

\subsection{Data preprocessing}

\subsubsection{Imbalanced Data Processing}\

\begin{figure}
    \centering
    \includegraphics[width=0.65\linewidth]{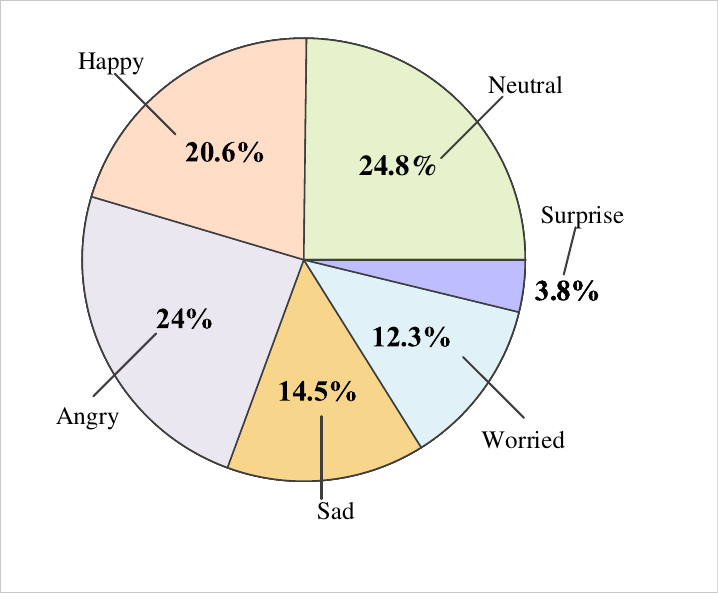}
    \caption{The distribution of MER2024-Train\&Val.}
    \label{fig1}
\end{figure}

The distribution of labeled data provided by MER2024-Train\&Val is shown in Figure~\ref{fig1}. Among them, the labels ``neutral'', ``angry'', and ``happy'' are dominant, ``worried" and ``sad'' are secondary, while ``surprise'' accounts for only 3.8\%. This dataset exhibits imbalance,
where minority class samples risk being overwhelmed by majority samples during training, potentially impacting model performance. To address this issue, we rebalance the training data distribution by employing random oversampling. 
Through this process, each emotion category achieves a predetermined sample count, resulting in a more balanced dataset.

\subsubsection{Noise embedding enhancement strategy}\ 

Various data augmentation techniques can be applied to the raw data across different modalities to enhance the generalization of the model. For instance, image flipping and random cropping can be utilized for visual data, while spectrogram transformation can augment audio data. These augmentation strategies help the model generalize better and perform more reliably in diverse scenarios.

However, integrating these diverse data enhancement methods with existing models can be challenging. To address this, inspired by the work of Wang et al. \cite{wang2023feenn, hazarika2022analyzing,fan2023learning, ho2020denoising}, we introduce a noise embedding enhancement (NEE) strategy to intervene in the representation of each modality. It constructs noise embeddings ${\{ {N^m}\} ^{m \in \{ v,a,t\} }}$ for each extracted feature ${\{ {F^m}\} ^{m \in \{ v,a,t\} }}$ using Gaussian noise. The noise schedule is built across time steps, and the process is described by the following formula:
\vspace{-2mm}
\begin{equation}
{N^m} = \sqrt {{\alpha _T}} F^m_0 + \sqrt {1 - {\alpha _T}} \epsilon
\vspace{-0.2cm}
\end{equation}
\begin{equation}
{\alpha _T} = \prod\limits_{t = 1}^T {(1 - {\beta _t})}
\vspace{-0.1cm}
\end{equation}
where $F^m_0$ denotes the embedding at time step 0, and $\epsilon$ is a noise vector randomly sampled from a Gaussian distribution, $\epsilon \sim {\mathcal{N}}\left( {0,1} \right)$. Timestep parameter $T$ is used to control the degree of noise injected into the data and simulate the process of gradual destruction of the data by random noise. The parameter ${\beta _t}$ is the $t$-th value of the schedule parameter sequence $\{ {\beta _1},{\beta _2}, \cdots ,{\beta _t}, \cdots ,{\beta _T}\}$. Here, the total number of steps $T$ affects the gradual transition of the original embedding to the noise-dominated state and is set to 100. In our experiments, ${\beta _1}$ and ${\beta _T}$ are set to 0.001 and 0.1, respectively. By embedding random noise at the feature vector level through the aforementioned NEE strategy, the model is encouraged to adapt to variations in sample data to enhance its generalization ability.

\begin{figure*}[t]
  \centering
  \includegraphics[scale=0.95]{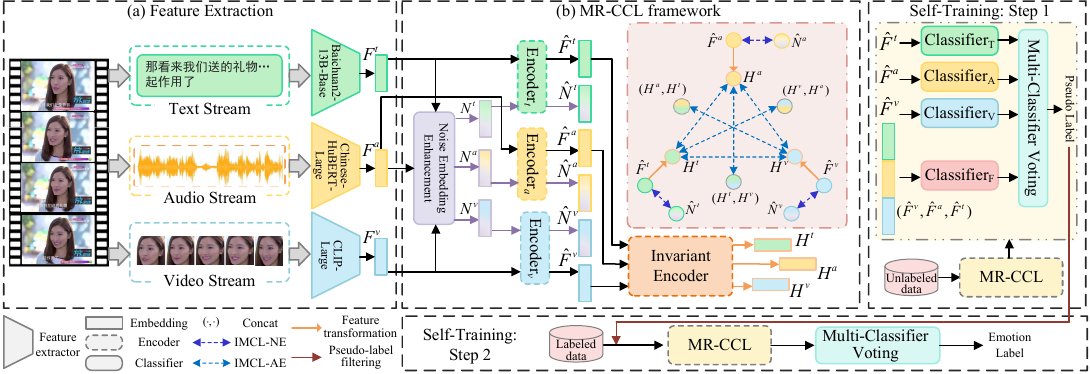}
  \caption{Illustration of our MR-CCL framework.}
  \label{fig2}
\end{figure*}

\subsection{Model Architecture}
\subsubsection{Modality Representation Combinatorial Contrastive Learning}\

\label{constractive}

We propose \textbf{M}odality \textbf{R}epresentation \textbf{C}ombinatorial \textbf{C}ontrastive  \textbf{L}earning (MR-CCL) framework to leverage the given unlabeled data more effectively. The overall architecture is illustrated in Figure \ref{fig2}.

In the realm of multimodal emotion recognition tasks, a proven strategy involves extracting features specific to each modality as well as invariant features that capture relationships between different modalities, which facilitates the fusion and interaction of diverse modal inputs \cite{zuo2023exploiting, liu2024contrastive}. To implement our framework, we begin by pre-training three specificity encoders and one invariant encoder using all available unlabeled data. This process is designed to enable the model to discern and encode the intrinsic characteristics and structural nuances of multimodal data. Each specificity encoder is composed of multiple Transformer layers, while the invariant encoder is structured with linear layers.

The input embedding ${\{ { F^m}\} ^{m \in \{ v,a,t\} }}$ for each modality along with their respective noise embeddings ${\{ { N^m}\} ^{m \in \{ v,a,t\} }}$ are fed into the specificity encoders to obtaining modality-specific embeddings ${\{ {\hat F^m}\} ^{m \in \{ v,a,t\} }}$ and ${\{ {\hat N^m}\} ^{m \in \{ v,a,t\} }}$. Subsequently, the invariant encoder transforms the ${\{ {\hat F^m}\} ^{m \in \{ v,a,t\} }}$ embeddings into a unified vector space, extracting modality-invariant emotion features ${\{ { H^m}\} ^{m \in \{ v,a,t\} }}$, a technique validated for its effectiveness in multimodal emotion recognition \cite{liu2024contrastive}. Following this, we proceed the combinatorial contrastive learning method based on the features obtained from the encoders above.

\textbf{(1) Intra-modality Contrastive Learning with Noisy Embedding (IMCL-NE).}  We employ contrastive learning techniques to learn a consistent representation between the modality-specific representation ${\{ {\hat F^m}\} ^{m \in \{ v,a,t\} }}$ and the noisy modality-specific representation ${\{ {\hat N^m}\} ^{m \in \{ v,a,t\} }}$ to enhanced feature robustness and model generalization performance. Noise contrastive estimation \cite{oord2018representation} is utilized to calculate intra-modality contrastive loss of each modality. Taking a pair of visual embeddings $\hat F^v$ and $\hat N^v$ as an example, it can be implemented through the following equation:
\vspace{-2mm}
\begin{align}
\begin{array}{c}
{L_{v\_fn}} =  - \frac{1}{B}\sum\limits_k^B {\log \left( {\frac{{\exp \left( {{{{{(\hat F_k^v)}^{\rm{T}}}\hat N_k^v} \mathord{\left/
 {\vphantom {{{{(\hat F_k^v)}^{\rm{T}}}\hat N_k^v} {{\tau _v}}}} \right.
 \kern-\nulldelimiterspace} {{\tau _v}}}} \right)}}{{\sum\nolimits_j^B {\exp \left( {{{{{(\hat F_k^v)}^{\rm{T}}}\hat N_j^v} \mathord{\left/
 {\vphantom {{{{(\hat F_k^v)}^{\rm{T}}}\hat N_j^v} {{\tau _v}}}} \right.
 \kern-\nulldelimiterspace} {{\tau _v}}}} \right)} }}} \right)} - \\
 - \frac{1}{B}\sum\limits_k^B {\log \left( {\frac{{\exp \left( {{{{{(\hat F_k^v)}^{\rm{T}}}\hat N_k^v} \mathord{\left/
 {\vphantom {{{{(\hat F_k^v)}^{\rm{T}}}\hat N_k^v} {{\tau _v}}}} \right.
 \kern-\nulldelimiterspace} {{\tau _v}}}} \right)}}{{\sum\nolimits_j^B {\exp \left( {{{{{(\hat F_j^v)}^{\rm{T}}}\hat N_k^v} \mathord{\left/
 {\vphantom {{{{(\hat F_j^v)}^{\rm{T}}}\hat N_k^v} {{\tau _v}}}} \right.
 \kern-\nulldelimiterspace} {{\tau _v}}}} \right)} }}} \right)}
\end{array}
  \label{eq3}
\end{align}
where $B$ is the batch size, $\tau_v$ is the temperature, ${( \cdot )^{\rm{T}}}$ represents the vector transpose operation, and ${L_{a\_fn}}$ and ${L_{t\_fn}}$ are calculated similarly. Thus, the total intra-modality contrastive loss ${L_{fn\_intra}}$ is given by: ${L_{fn\_intra}} = {{({L_{v\_fn}} + {L_{a\_fn}} + {L_{t\_fn}})} \mathord{\left/
 {\vphantom {{({L_{v\_fn}} + {L_{a\_fn}} + {L_{t\_fn}})} 3}} \right.
 \kern-\nulldelimiterspace} 3}$.

\textbf{(2) Inter-Modality Contrastive Learning with Aligned Embeddings (IMCL-AE).} 
Inspired by Shvetsova et al. \cite{shvetsova2022everything}, we aim to calculate the inter-modality contrastive loss between all possible modality combinations, mapping the inputs from single or multiple modalities into a joint embedding space where semantically similar inputs are closely positioned. First, the contrastive losses between single modalities are calculated using three pairs of embeddings: $({{H}^v}, {{H}^a})$, $({{H}^a}, {{H}^t})$, and $({{H}^v}, {{H}^t})$, from which the losses $L_{av}$, $L_{at}$, and $L_{vt}$ are respectively calculated according to Equation (\ref{eq3}). Subsequently, we learn the close representation between one modality and the other two modalities. For instance, considering the embedding $H^v$ and concatenated vector $H^{at}= {\rm Concat}(H^a, H^t)$, this relationship can be quantified through the following equation:
\begin{align}
\vspace{-2mm}
\begin{array}{c}
{L_{v\_at}} =  - \frac{1}{B}\sum\limits_k^B {\log \left( {\frac{{\exp \left( {{{{{(H_k^v)}^{\rm{T}}}H_k^{at}} \mathord{\left/
 {\vphantom {{{{(H_k^v)}^{\rm{T}}}H_k^{at}} {{\tau _{at}}}}} \right.
 \kern-\nulldelimiterspace} {{\tau _{at}}}}} \right)}}{{\sum\nolimits_j^B {\exp \left( {{{{{(H_k^v)}^{\rm{T}}}H_j^{at}} \mathord{\left/
 {\vphantom {{{{(H_k^v)}^{\rm{T}}}H_j^{at}} {{\tau _v}}}} \right.
 \kern-\nulldelimiterspace} {{\tau _{at}}}}} \right)} }}} \right)} - \\
 - \frac{1}{B}\sum\limits_k^B {\log \left( {\frac{{\exp \left( {{{{{(H_k^v)}^{\rm{T}}}H_k^{at}} \mathord{\left/
 {\vphantom {{{{(H_k^v)}^{\rm{T}}}H_k^{at}} {{\tau _{at}}}}} \right.
 \kern-\nulldelimiterspace} {{\tau _{at}}}}} \right)}}{{\sum\nolimits_j^B {\exp \left( {{{{{(H_j^v)}^{\rm{T}}}H_k^{at}} \mathord{\left/
 {\vphantom {{{{(H_j^v)}^{\rm{T}}}H_k^{at}} {{\tau _{at}}}}} \right.
 \kern-\nulldelimiterspace} {{\tau _{at}}}}} \right)} }}} \right)}
\end{array}
  \label{eq4}
\end{align}
where $\tau_{at}$ is the temperature, \textit{B} is batch size, ${L_{a\_vt}}$ and ${L_{t\_av}}$ are calculated similarly. The total inter-modality contrastive loss ${L_{imc}}$ is given by: ${L_{imc}} = {{({L_{av}} + {L_{vt}} + {L_{at}} + {L_{v\_at}} + {L_{a\_vt}} + {L_{t\_av}})} \mathord{\left/
 {\vphantom {{({L_{av}} + {L_{vt}} + {L_{at}} + {L_{v\_at}} + {L_{a\_vt}} + {L_{t\_av}})} 6}} \right.
 \kern-\nulldelimiterspace} 6}$.

 \subsubsection{Self-Training Strategy}\

Pseudo-labeling plays a significant role in utilizing unlabeled data. According to previous works \cite{li2023mining}, different models have different preferences for classification. Similarly, we observe that the baseline model performs better on the ``happy'', ``worried'', and ``surprise'' categories, while our model performs better on the other categories. So we take the intersection between the inference results of these two models, generate and filter pseudo labels according to a certain confidence threshold.

Firstly, we train an initial model and a baseline model using existing labeled data within our contrastive learning framework described in Section \ref{constractive}. These models are subsequently employed to predict unlabeled data and generate pseudo-labels. Next, we apply a confidence threshold to select predictions with high confidence. These high-confidence pseudo-labeled samples are incorporated into the training set as additional labeled data. Finally, the model is retrained using the augmented training set. Note that all the pseudo-labeled samples will not be divided into the validating set.

However, the pseudo-labeling strategy also has potential drawbacks, particularly when incorrect pseudo-labels introduce erroneous information into the model. To mitigate these risks, we separate the training process into two steps. In the first step, we refrain from incorporating any pseudo-labeled samples, which aims to establish a stable and relatively accurate model from the outset. In the second step, we introduce the pseudo-labeled samples into the training set while reducing the learning rate to gradually improve the model and decrease the influence of wrong labels.

\vspace{-0.15cm}
\subsubsection{Multi-Classifier Voting}\ 


We utilize a weighted soft voting strategy for classification, which combines outputs from four classifiers: one modality fusion classifier (Classifier F) and three unimodal classifiers (Classifiers A, V, and T). The unimodal classifiers receive modality-specific features and independently predict emotion categories. The modality fusion classifier processes a joint representation of multimodal inputs and performs emotion classification tasks.

The process begins with aggregating confidence scores from the four classifiers and assigning specific weights to each result. Subsequently, these weighted confidence scores are compared, and the classification with the highest confidence is selected.
According to prior findings from the MER2024 Baseline \cite{2024mer}, the audio modality classifier (Classifier A) has shown the highest individual accuracy. Therefore, Classifier A is assigned the highest weight. Following this, Classifiers V and T receive weights in descending order. The fusion classifier (Classifier F) is assigned a weight equal to Classifier A's weight to capitalize on its multimodal information.


\vspace{-0.1cm}
\section{EXPERIMENTS AND RESULTS}
\subsection{Datasets and Evaluation Metric}\ 

MER2024 \cite{lian2023mer} is a Chinese emotion dataset designed for emotion recognition challenge, comprising four subsets: Train\&Val, MER-SEMI, MER-NOISE, and MER-OV. The Train\&Val subset consists of 5030 labeled single-speaker video segments used for training and validation purposes in the MER2024 challenge. The MER-SEMI subset contains a total of 115,595 unlabeled data, with 1169 of these serving as the true test set for evaluating performance in the Track1 `semi-supervised learning challenge'. Participants in this track are tasked with predicting discrete emotions through a large number of unlabeled samples. They are encouraged to employ semi-supervised learning techniques to improve emotion recognition performance. Here, discrete emotion labels include 6 classes, i.e., worried, happy, neutral, angry, surprised, and sad. Similar to the baseline \cite{2024mer}, we adopt weighted average F-score (WAF) \cite{lian2021ctnet} to evaluate the overall recognition performance due to the inherent class imbalance.

\vspace{-0.25cm}
\subsection{Implementation Details}\ 

We conduct all experiments on an NVIDIA A100 80GB GPU. For feature extraction, 
the embedding dimensions ${d_v}$, ${d_a}$, and ${d_t}$ for video, audio, and text are 768, 1024, and 5120, respectively.

In the process of pseudo label selection, we apply a threshold of 0.99 for the categories ``happy'', ``neutral'', ``angry'', and ``sad'' to determine pseudo labels. Given the inherent imbalance in the categories ``worried" and ``surprise", we use a lower threshold of 0.85. Table \ref{tab: pseudo label} presents the specific counts of pseudo labels assigned to each class. The weight of classifiers A, V, T, and F are set to [0.7, 0.5, 0.4, 0.7]. During training, we employ the Adam optimizer for MR-CCL, setting the initial learning rate to 1e-4 and utilizing a batch size of 512. We train the model with a maximum of 40 epochs and employ an early stopping strategy with a patience of 5 epochs. 

In the final training process, we incorporate both pseudo-label and oversample strategies. In the first step, we use the oversampling strategy to increase the three minority classes to 850 samples; in the second step, we add the data with false labels to the training set, increase the number of oversampling to 1000, and reduce the learning rate from 1e-4 to 5e-5. Each step contains 20 epochs.

\vspace{-0.26cm}
\subsection{Main Results and Analysis}\ 

Based on the MER-SEMI track theme, we employ two main ways to utilize unlabeled data: 1) The contrastive learning method based on unlabeled data, and 2) Expanding the training set by labeling unlabeled data with pseudo labels.
Table \ref{tab:commands} summarizes the results of our model in comparison to the baseline, as well as the outcomes from ablation experiments. The results indicate that using unlabeled data significantly enhances multimodal emotion recognition performance.

We present the results of expanding the training set with only the pseudo-labeling strategy or the oversample strategy in Table \ref{tab: pseudo label} and Table \ref{tab: oversample}. We can observe that utilizing the pseudo-label strategy gets a higher WAF on the validating set (and is more closer to the testing set) than using the original training set and the oversample strategy. This phenomenon proves that the pseudo-labeled data provides more diverse training data, improves the generalization ability of the model, and may make the data distribution of the training set closer to the data distribution of the test set, thus reducing the impact of distribution differences on the model performance. In addition, with the addition of pseudo-labeled data, the overfitting phenomenon of the model has been alleviated, and the generalization ability and performance of the model can be improved.
We can also observe that the result of the pseudo-label strategy on the test set is similar to the oversample strategy. This may be because the distribution of the testing set still has differences from the validating and training sets, or because the model is overconfident in its predictions and generates false pseudo labels. Therefore, there is still room for improvement in the pseudo-labeling strategy. Table \ref{lr} also shows that the decrease in learning rate helps stabilize the performance of the model, which provides support for the above analysis.

\begin{table}
  \caption{Main results on the train\&val and test set.}
  \label{tab:commands}
\begin{tabular}{lcc}
\toprule
\multirow{2}{*}{Systems}    & Train\&Val & MER-SEMI \\
                            & WAF(↑)    & WAF(↑)   \\ \midrule
Baseline                    & 79.31     & 86.73    \\
\textbf{Ours}                        & 86.53    & \textbf{88.25}   \\
w/o contrastive pre-training        & 85.72    & 85.86   \\
w/o NEE strategy                    & 86.02    & 87.64   \\
w/o pseudo labels & 80.40    & 86.35   \\
w/o oversample & 85.17    & 86.07   \\
\bottomrule
\end{tabular}
\end{table}

\begin{table}[]

\caption{The data distribution and WAF accuracy with pseudo labels. The data ``1418(+170)'' means that 170 ``neutral'' samples with pseudo labels are added to the training set, which has 1418 ``neutral'' samples in total.}
\label{tab: pseudo label}
\resizebox{1.0\linewidth}{!}{
\begin{tabular}{lcc}
\toprule
Emotion   & Train\&Val & Pseudo label \\
\midrule
Neutral  & 1248     & 1418(+170)         \\
Angry    & 1208     & 1330(+122)         \\
Happy    & 1038     & 1244(+206)         \\
Sad      & 730      & 1159(+429)         \\
Worried  & 616      & 937(+321)          \\
Surprise & 190      & 340(+150)          \\
Total    & 5030     & 6428(+1398)  \\ \midrule
Train\&Val (WAF ↑)    & 79.31      & \textbf{88.73}       \\
MER-SEMI (WAF ↑)      & 86.73       & 87.12 \\ 
\bottomrule       
\end{tabular}
}
\end{table}

\begin{table}[]
\caption{The impact of oversampling several minority classes on the training set. The mark ``$*$'' represents the baseline data. The numbers in parentheses represent how many duplicated samples were added to the class compared to the baseline.}
\label{tab: oversample}
\resizebox{1.0\linewidth}{!}{
\begin{tabular}{ccccc}
\toprule
\multirow{2}{*}{Sad} & \multirow{2}{*}{Worried} & \multirow{2}{*}{Surprise} & Train\&Val & MER-SEMI \\
                     &                          &                           & WAF(↑)     & WAF(↑)   \\ \midrule
                     730$^*$                  & 616$^*$                      & 190$^*$                       & 79.31$^*$      & 86.73$^*$    \\ \midrule
                     \multicolumn{5}{c}{Oversample Strategy} \\
                     \midrule

850(+120)           & 850(+234)                & 850(+660)                & 83.03      & 88.05   \\
900(+170)                    & 900(+286)                        & 900(+710)                         & 81.50          & 86.57        \\
800(+70)                    & 800(+284)                        & 800(+610)                         & 82.14          & 87.47       \\
850(+120)                    & 616(+0)                        & 190(+0)                         & 80.18          & 86.20        \\
730(+0)                    & 850(+234)                        & 190(+0)                         & 79.84          & 85.36         \\ 
730(+0)                    & 616(+0)                        & 850(+660)                        & 81.38          & 86.11         
\\ \bottomrule  
\end{tabular}
}
\end{table}

\begin{table}[]
\caption{The impact of learning rate adjustment strategy.}
\label{lr}
\centering
\resizebox{0.85\linewidth}{!}{
\begin{tabular}{lcc}
\toprule
\multirow{2}{*}{Learning Rate} & Train\&Val     & MER-SEMI       \\
                               & WAF(↑)         & WAF(↑)         \\ \midrule
1e-4(Unchanged)                & 85.42          & 87.33          \\
5e-5(Reduced)                  & \textbf{86.53} & \textbf{88.25} \\ \bottomrule
\end{tabular}
}
\end{table}

\section{CONCLUSION}\ 

This paper describes our proposed semi-supervised multimodal emotion recognition method for the MER2024-SEMI challenge. Our method comprises three main steps: modality representation combinatorial contrastive learning, self-training, and multi-classifier voting. Additionally, we perform oversampling to address the class imbalance problem. The effectiveness of the proposed method is validated, achieving a weighted average F-score of 88.25\% on the test set of the MER2024-SEMI challenge. We intend to consider the stronger semi-supervised training strategy and fusion module in future work.

\section{Acknowledgments}
The research by Miao Ma is funded by the National Natural Science Foundation of China under Grant 62377031, and the Key Research and Development Program in Shaanxi Province under Grant 2023-YBGY-241. The work is also supported by the following fundings: National Natural Science Foundation of China (62066033), the Outstanding Youth Foundation of the Natural Science Foundation of Inner Mongolia (2022JQ05), the Science and Technology Program of Inner Mongolia Autonomous Region (2021GG0158), the Hohhot Collaborative Innovation Project for Universities and Institutes, the Youth Science and Technology Talent Cultivation Project of Inner Mongolia University (21221505), the fund of Supporting the Reform and Development of Local Universities (Disciplinary Construction) and the special research project of First-class Discipline of Inner Mongolia A. R. of China under Grant (YLXKZX-ND-036).
College of Computer Science, Inner Mongolia University, National \& Local Joint Engineering Research Center of Intelligent Information Processing Technology for Mongolian, and Inner Mongolia Key Laboratory of Multilingual Artificial Intelligence Technology provided support to this work.

\bibliography{aaai25}

\end{document}